\documentclass[10pt]{iopart}

\usepackage{iopams}
\usepackage{graphicx}
\usepackage{colordvi,color}
\usepackage{cite,etoolbox}
\usepackage[normalem]{ulem}
\usepackage[dvipsnames]{xcolor}

\begin{document}

\title[Data-driven mass-spring Calliphora wing model]{An experimental data-driven mass-spring model of flexible \emph{Calliphora} wings}

\author{Hung Truong$^{1}$, Thomas Engels$^{2}$, Henja Wehmann$^{2}$, \\ 
        Dmitry Kolomenskiy$^{3,4}$, Fritz-Olaf Lehmann$^{2}$, \\
        Kai Schneider$^{1}$}

\address{$^{1}$ Aix-Marseille University, CNRS, I2M, Marseille, France.}
\address{$^{2}$ Department of Animal Physiology, University of Rostock, Rostock, Germany.}
\address{$^{3}$ Tokyo Institute of Technology, Tokyo, Japan.}
\address{$^{4}$ Present address: Center for Design, Manufacturing and Materials, Skolkovo Institute of Science and Technology, Moscow, Russia.}

\ead{dinh-hung.truong@univ-amu.fr}
\vspace{10pt}
\begin{indented}
\item[]October 2021
\end{indented}

\begin{abstract}
Insect wings can undergo significant deformation during flapping motion owing to inertial, elastic and aerodynamic forces. Changes in shape then alter aerodynamic forces, resulting in a fully coupled Fluid--Structure Interaction (FSI) problem. Here, we present detailed three-dimensional FSI simulations of deformable blowfly (\textit{Calliphora vomitoria}) wings in flapping flight. A wing model is proposed using a multi-parameter mass-spring approach, chosen for its implementation simplicity and computational efficiency. We train the model to reproduce static elasticity measurements by optimizing its parameters using a genetic algorithm with covariance matrix adaptation (CMA-ES). Wing models trained with experimental data are then coupled to a high-performance flow solver run on massively parallel supercomputers. Different features of the modeling approach and the intra-species variability of elastic properties are discussed. We found that individuals with different wing stiffness exhibit similar aerodynamic properties characterized by dimensionless forces and power at the same Reynolds number. We further study the influence of wing flexibility by comparing between the flexible wings and their rigid counterparts. Under equal prescribed kinematic conditions for rigid and flexible wings, wing flexibility improves lift-to-drag ratio as well as lift-to-power ratio and reduces peak force observed during wing rotation.
\end{abstract}

%
\vspace{2pc}
\ioptwocol

\section{Introduction}

The wings of an insect are hundreds of times lighter than its body, yet they sustain dynamic loads that exceed the body weight. Consequently, they deform significantly during flapping flight. To deal with these large deformations, insects have evolved highly compliant wings from which they benefit in many aspects: enhanced aerodynamic efficiency \cite{FSINakata2012, Zheng_2013_plosone}, flight stability \cite{Mistick_2016_jeb}, enhanced flight control \cite{WingDeformWalker3}, damage resistance \cite{Mountcastle_2014_jeb}, robustness to collisions \cite{Phan_2020_science}, to name a few.

For understanding the aerodynamics of insect locomotion, the fluid–structure interaction problem must be addressed by coupling fluid with solid mechanics. Computational methods yield insight into the instantaneous flow field surrounding the studied insect and with access to all aerodynamic quantities, which are difficult to obtain in experiments. Thus fundamental mechanisms behind the nonlinear dynamics of the flow can be revealed. However, numerical studies of insect flight are not trivial due to their high complexity. For simplification, studies are usually employing either completely rigid wings (e.g., \cite{HLiuRigidwings98,HLiuRigidwings09,YokoyamaRigidwings07}), or prescribed time-varying deformation \cite{Du,Zheng_2013_plosone}. Fully coupled fluid-structure interaction simulations of flapping insect wings are still challenging and give controversial results. While some studies found advantages of wing flexibility on aerodynamic performance of insects \cite{FSINakata2012,FSINakata2011,WingDeformMountcastle,Combesflexbb,FlexInfluenceTobing}, others reported negative impact on lift production \cite{FlexInfluenceZhao,FSITanaka2011,FlexInfluenceMeerendonk}. The anisotropy and inhomogeneity of the elastic properties of wings are clearly important factors in these studies. 

During flight, the architecture and material properties of insect wings determine predominantly their deformations, which are mostly passive \cite{CombesII}. Therefore, determining wing stiffness is critical to the modeling of insect wing dynamics~\cite{Ramananarivo_2011_pnas,HungRevol}. In combination with other functional requirements such as wing folding, hemolymph transport, etc.,  evolution has led to complex designs with individual shapes and sizes of veins, different types of hinges, resilin patches and varying thickness of the membrane. As a matter of fact, Young's modulus of insect wings may change from tens to hundreds of \si{\mega\pascal} between species or even different parts of the wing \cite{HenjaCalLocalStiffness}. Measurement conditions play also a crucial role in determining the wing stiffness due to wing desiccation. Altogether, the distribution of flexural rigidity in insect wings and its effects on wing dynamics are still poorly known.

In the past, only few numerical studies took into consideration these complex structures of wings. Combes and Daniel~\cite{CombesI} measured the overall flexural stiffness $EI$ either in spanwise or chordwise directions by assuming that wings were homogeneous beams. The data were then used in a simplified finite element model of a \textit{Manduca} wing. Nakata and Liu~\cite{FSINakata2012} and Tobing et al.~\cite{FlexInfluenceTobing} also set the parameters for their flexible wing models based on the measurements of Combes and Daniel. Nakata and Liu proposed an anisotropic hawkmoth wing model. On the other hand, Tobing et al. considered a 3D flexible wing model of bumblebees with uniform and reduced-tip stiffness. Ishihara et al.~\cite{FlexInfluenceIshihara} employed a model composed of a rigid leading edge connected with a rigid plate through a plate spring. The torsional stiffness of the latter was defined based on dynamic similarity. Nguyen et al.~\cite{FlexInfluenceNguyen} modeled a fruit fly wing where sharp variations in material properties of stiff veins and soft adjacent membrane were taken into account. In our previous work~\cite{HungRevol,HungFlappingBB_TAML}, numerical simulations of bumblebees with flexible wings using the open source framework FLUSI \cite{Flusi} have been carried out. We studied the impact of wing flexibility on aerodynamic forces by varying the Young's modulus $E$ of the vein system in a chosen range. Even though these contributions succeeded to include the venation structure in their wing models, the identification of the flexural rigidity for both the veins and the membranes remains a daunting task.

In the present paper, we propose a numerical method for estimating wing stiffness. It consists of a relatively simple mass-spring model based on the wing planform and venation pattern measured from photographs of the blowfly ({\it Calliphora vomitoria}). The stiffness parameters are now optimized considering acquired experimental data of real insect wings using a genetic algorithm with covariance matrix adaptation strategy. This approach ensures that the model has the same behavior as the real wing specimens in static bending tests. The wing model with the optimized stiffness is then used for three-dimensional unsteady FSI simulations of flapping wings at high resolution on massively parallel computers.

The remainder of the manuscript is organized as follows. In the material and methods section~\ref{MaterialsMethods}, we describe the wing model, the experimental as well as the mathematical methods that are used. The numerical results are then discussed in section~\ref{ResultsDiscussion}, starting with the validation of the optimization. Finally, conclusions of the study are drawn in section~\ref{section:conclusion} and some perspectives on future work are given.

\section{Materials and Methods\label{MaterialsMethods}}

\begin{figure*}
\centering
  \includegraphics[width=152.75mm]{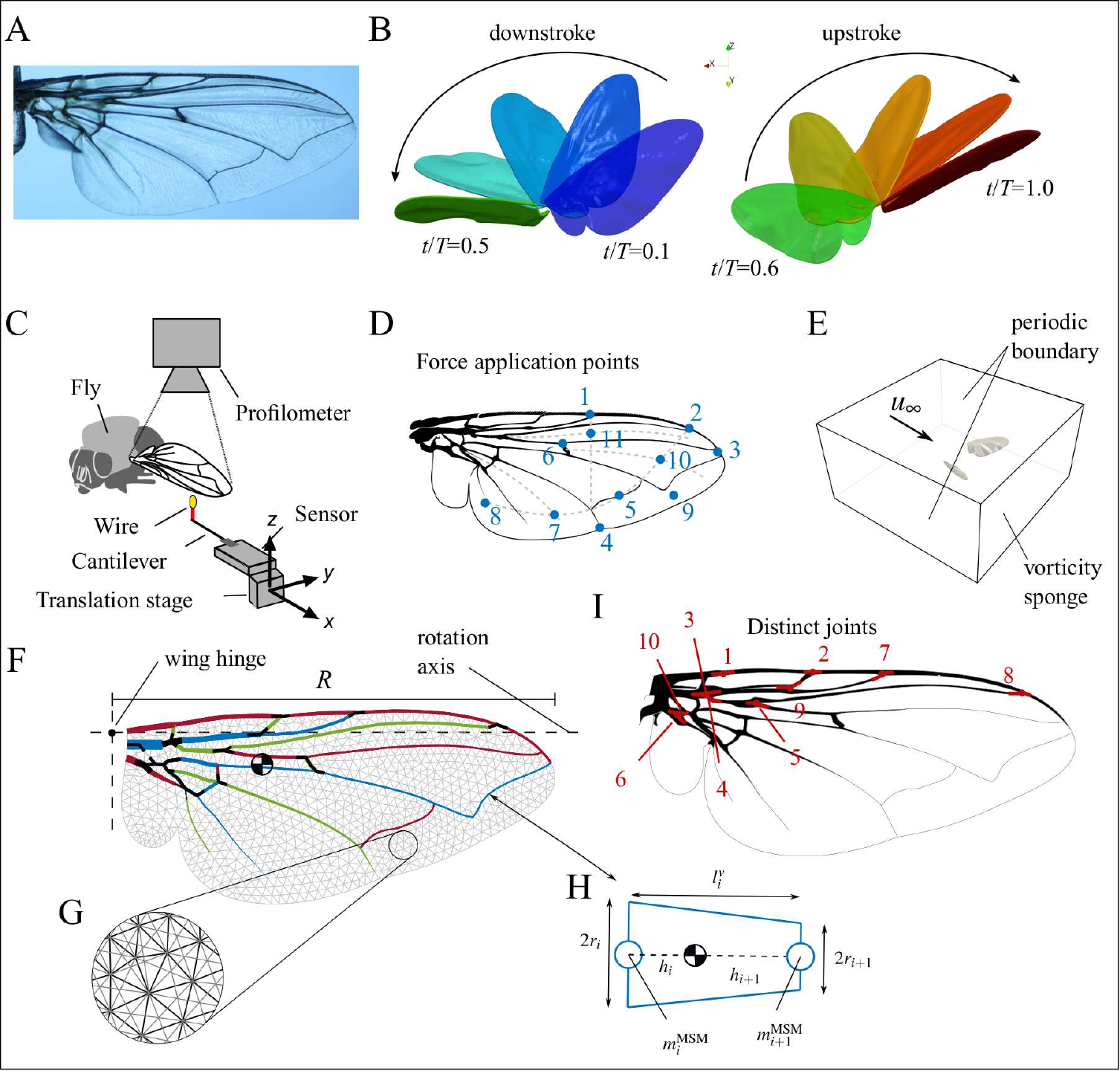}
  \caption{Morphology and modeling of wings. (A) Photograph of a blowfly wing. (B) Wing beat kinematics, taken from \cite{ThomasCalWingCorrugation} during up- and downstroke. (C) Experimental setup used in \cite{HenjaCalLocalStiffness} for force/deformation measurements. A point force is applied at one of the locations shown in (D) and the deformed wing surface is measured using an optical profilometer. (E) Sketch of the setup for fluid-structure interaction simulations. We simulate two flapping wings exposed to a mean flow of $u_\infty=1.35$ m/s. (F) Computational mesh used in the mass-spring model. Veins are colored for segment distinction, membrane mesh (extension springs) shown in gray. Vein diameters shown to scale. Black/white marker is the center of mass. (G) Zoom on the membrane mesh, showing cross-springs. (H) A segment of a vein between two mass points is modeled as conical frustum with varying diameter (I) Joints on the wing considered in optimization problem.\label{fig1meta}}
\end{figure*}

\subsection{Morphology and Wingbeat Kinematics of Blowfly Wings\label{sec:flex}}

An experimental setup for measuring the elasticity of female blowfly wings \emph{(Calliphora vomitoria)} was designed by Wehmann et al.~\cite{HenjaCalLocalStiffness} and the measured data are used in the present work. As details on the measurements can be found in \cite{HenjaCalLocalStiffness}, we limit the presentation here to a minimum. Wing shape and venation structure are extracted from high resolution photographs of blowfly wings (Fig.~\ref{fig1meta}A), taken using a camera (EOS-750D, Canon) attached to a stereomicroscope (Stemi 508, Zeiss) with a resolution of \SI{0.46}{\micro\metre}/pixel. This high resolution allows us to determine vein diameters, assuming circular cross section. In reality, veins are hollow tubes filled with hemolymph which supplies nutrients and other factors to the wing's living tissues \cite{PassVeinCirculatory}. However, this structure is too complex to be taken into account by our model. Consequently, veins are considered as solid rods of cuticle with density $\rho_c =$ \SI{ 1300}{\kilo\gram\per\cubic\metre} as given in~\cite{CuticleProperties}. This assumption has the tendency to overestimate vein mass. Ganguli et al.~\cite{RanCalwingmass} weighed 10 wing pairs and found their individual mass between \SI{200.3}{\micro\gram} and \SI{272.3}{\micro\gram} with a standard deviation of \SI{22.94}{\micro\gram}. For our model, total wing mass is set to \SI{250}{\micro\gram}~\cite{RanCalwingmass}. The numerical wing length $R$, averaged over nine wing models used in our study, is \SI{9.01 \pm 0.18}{\milli\metre}. This accounts for a relative difference of $3\%$ compared to the measured wing length \SI{9.29 \pm 0.20}{\milli\metre} given in \cite{HenjaCalLocalStiffness}. The discrepancy can be explained by the fact that the wing model was built based on detached wings and parts of the wing roots were omitted.

Wing beat kinematics vary between individual flies of the species considered as well as between cycles in the same individual. There is hence no `true' wing beat kinematics for a fly. It is therefore appropriate (cf. the discussion in \cite{ThomasCalWingCorrugation}, suppl. mat.) to use the generic wing kinematic protocol proposed in \cite{ThomasCalWingCorrugation}. The wing beat is visualized in Fig. \ref{fig1meta}B. In tethered flight, mean stroke frequency $f$ in blowflies varies between $127$ and \SI{180}{\Hz} with a mean value \SI{158}{\Hz} \cite{NachtigallCalfreq}. Hence, the stroke frequency $f=$ \SI{158}{\Hz} and the stroke amplitude $\Phi = 135^{\circ}$ are used in our study.

\subsection{Measurements of Wing Elasticity\label{sec:measurements}}
 The experimental setup used for measuring local deformation of blowfly wings under external loads is illustrated in Fig.~\ref{fig1meta}C. A living blowfly was mounted on a holder and during the measurements, wings were kept attached to the living body to limit dry-out effects. Otherwise, insect wings would quickly stiffen and become brittle \cite{PassVeinCirculatory, HenjaCalLocalStiffness}. Point forces were applied at different locations (Fig. \ref{fig1meta}D) and with different magnitudes. Force magnitude was measured by a small, cantilever force sensor. The surface of wings with and without force application was measured using an optical profilometer that projected a grid on the wing surface and recorded local vertical height $z(x,y)$ with a resolution of $384 \times 512$ points in $x$ and $y$ direction. These data are thus specified in an Eulerian reference frame. The data from these experiments are used as reference data for training of our numerical model (see below).

\subsection{Mass-Spring Model for Elastic Wings}
The complex compound structure of insect wings poses significant challenges to mechanical modeling. Our model must reproduce experimental data while being computationally efficient and straightforward in its implementation. We choose a mass-spring approach and hence model the wing by a system of discrete mass points connected by linear extension and bending springs. The training procedure for parameter identification could in principle identify the underlying mechanical structure starting from randomly distributed mass points and spring coefficients~\cite{MSMLloyd,MSMBianchiParam}. However, the training can be greatly accelerated by using an initial configuration that takes the distinct properties of veins and membrane into account. This functional approach \cite{HungRevol} is chosen here, resulting in the static computational mesh (Fig.~\ref{fig1meta}F). This distribution of mass points as well as their connectivity via the springs is thus predetermined. Veins are composed of bending- and extension springs, the latter being set to a large value $k^{e}_v=\SI{1000}{\newton\per\centi\metre}$ to approximate an inextensible vein. Note that veins composed of perfectly rigid segments result in a globally coupled numerical problem with increased computational cost. The membrane is modeled using only extension springs, where the spring constant is set to $k^{e}_m=\SI{1000}{\newton\per\centi\metre}$ (Fig. \ref{fig1meta}F). While an idealized membrane resists only to stretching and not to bending, real membrane tissue does have a slight bending resistance. Hence, we add `cross-springs' (Fig. \ref{fig1meta}G) to give the membrane a slight bending stiffness~\cite{HungFlappingBB_TAML}. This latter suppresses artifacts at the trailing edge of the wing where no discernible vein supports the membrane. Both types of extension springs in the membrane are set to the same stiffness value $k^{e}_m$ in order to preserve isotropic behavior under stretching~\cite{HungRevol}.

Besides the veins and the membrane, wings contain multiple discrete joints where the vein is either interrupted or abruptly changes in stiffness. These joints were identified by manipulating wings under a microscope. There are several joints in a blowfly wing but only 10 joints (Fig. \ref{fig1meta}I) are included in the training problem. A sensitivity analysis, determining how much changes in the joint stiffness parameters alter the deformation, had been performed prior to the optimization runs. This helped us to decide which joints should be included in the training problem.

\subsection{Mass Distribution\label{subsec:Calliphora_mass_distribution}} 
Since our model is trained using static bending tests and because wings are lightweight (and consequently, the effect of gravity is negligible), the mass distribution does not enter the training problem and is therefore determined \emph{a priori}. 

Cross sections of veins play an important role in the identification of wing elastic properties. They allow us to determine the vein volume and the vein second moments of area $I$. However, realistic shapes of vein sections are varying from wing root to wing tip and these data are currently unavailable~\cite{CicadaVein}. In our model, the variation of vein diameter along wingspan is taken into account by assuming each vein segment as a conical frustum. 
The diameters at both ends of each segment are determined from photographs (see Supplementary materials, section 2). The distribution of the mass onto the limiting discrete mass points is then calculated based on this assumption. We consider a segment $i$ of a vein as shown in Fig.~\ref{fig1meta}H. The radii are $r_i$ and $r_{i+1}$, where $r_i > r_{i+1}$. The mass of the segment $m_i^v$ is distributed into the two mass points $m_i^\mathrm{MSM}$ and $m_{i+1}^\mathrm{MSM}$ such that the two systems have the same center of mass. This condition is satisfied by the relation:
\begin{equation}
    h_i m_{i}^v = m_{i+1}^\mathrm{MSM} (h_i + h_{i+1}) = m_{i+1}^\mathrm{MSM} l_i^v,
\label{eqn:CG_vein_mass_distribution}
\end{equation}
where $h_i$ is the centroid of a conical frustum,
\begin{equation}
    h_i = \frac{l_i^v}{4} \frac{(r_i^2 + 2 r_i r_{i+1} + 3r_{i+1}^2)}{(r_i^2 + r_i r_{i+1} + r_{i+1}^2)}.
\label{eqn:centroid_conical_frustum}
\end{equation}

Combining eqns.~(\ref{eqn:CG_vein_mass_distribution}) and (\ref{eqn:centroid_conical_frustum}) yields $m_{i+1}^\mathrm{MSM}  = m_{i}^v h_i$ and $m_{i}^\mathrm{MSM}  = m_{i}^v - m_{i+1}^\mathrm{MSM}$, which is applied to all veins. The vein system accounts for $67.41\%$ of the total wing mass, the remainder is attributed to the membrane. In bumblebee wings, we previously used the measured center of gravity to fit a bi-linear mass distribution \cite{BumblebeeWingStructure}. This results in the membrane thickness tapering off toward tip and trailing edge. Such measurements are not available for blowflies, thus we used the same bi-linear function, scaled using wing mass and wing length of blowflies. This yields the mass of the $i^{th}$ point on the membrane:
\begin{equation}
m_i = (1.75  - 1.47 x_i + 1.01 y_i )\cdot 10^{-7}
\end{equation}
where $m_i$ is the mass calculated in \si{\kilo\gram}, $x_i$ and $y_i$ are the distances in \si{\centi\metre} from the $i^{th}$ mass point to the wing root and the rotation axis (cf. Fig.~\ref{fig1meta}F), respectively.

\subsection{Training of Wing Model with Experimental Data \label{section:genetic_optimization}} 
Besides the mass distribution, the mass-spring model for the flexible wing features the stiffness distribution to be determined. In reality, veins are hollow and the shape of the vein cross section clearly plays an important role in estimating the wing stiffness. However, the available published data on this subject are still sparse~\cite{CicadaVein}. For simplification, we approximate all veins as having solid circular section of which the second moment of area can be determined from the diameter, $I \propto d^4$. Since the bending rigidity is a product of the Young's modulus $E$ and the moment of area $I$, a numerical optimization is used to adjust the Young's modulus such that the vein model deforms by the same amount as the real hollow non-circular veins in static bending tests. In addition, insect wings contain a number of joints, the flexibility of which depends on several factors \cite{WeisVeinJointStiffness}. Consequently, the flexibility of these joints cannot be estimated based on the flexural rigidity $EI$ of veins and needs to be optimized. However, taking into account all joints in the training process is expensive. In order to reduce the number of optimized parameters, numerical sensitivity tests on joints were performed and those having little impact on the wing deformation were not be included in the optimization. This results in a total of 10 joints (Fig.~\ref{fig1meta}I) needed to be optimized. The membrane is essentially an inextensible sheet with small bending stiffness (modeled using `cross-springs', Fig.~\ref{fig1meta}G) and is not included in the training. Thus, a set $\kappa \in \mathbb{R}^n_+$ with $n=11$ parameters are to be determined from experimental data.

Measuring dynamic wing deformation at high frequencies during flight is challenging. Thus, most qualitative data on wing deformation are based on photographs and not time-resolved \cite{WingDeformCombesStacey}. Even though some studies succeed in recording the wing surface deformation during wing flapping motion \cite{WingDeformMountcastle,WingDeformKoehler}, data on external inertial and aerodynamic forces acting on the wing are inaccessible. As an alternative, we use static measurements of wing deformation under known loads. As long as local deformation is not large enough to cause non-linear bending behavior, non-linearity in the model stems only from large deflections (geometric nonlinearity, \cite{ENGELS2015JCP}) and interactions with the fluid. This is usually the case in insect wings and consequently enables using static measurements for training.

In general, for each of the nine individual wings, a number of $N_\mathrm{exp}$ measurements was performed with different force magnitudes and application points (Fig.~\ref{fig1meta}D), scoring the complete wing surface in the deformed state. Some data were too noisy and omitted from the set, thus $8 \leq N_\mathrm{exp} \leq 11$. For training of the numerical model, each of these trials is repeated numerically with the current set of parameters $\kappa$. Damping coefficients are added to the solid model to achieve a steady state, because fluid is excluded in these simulations. Then, the quality of $\kappa$ is assessed by evaluating a cost function to quantify the error, which is subsequently used to update $\kappa$.

\subsubsection{Definition of cost function:}
\begin{figure}
\centering
\fbox{
\includegraphics[width=0.95\linewidth]{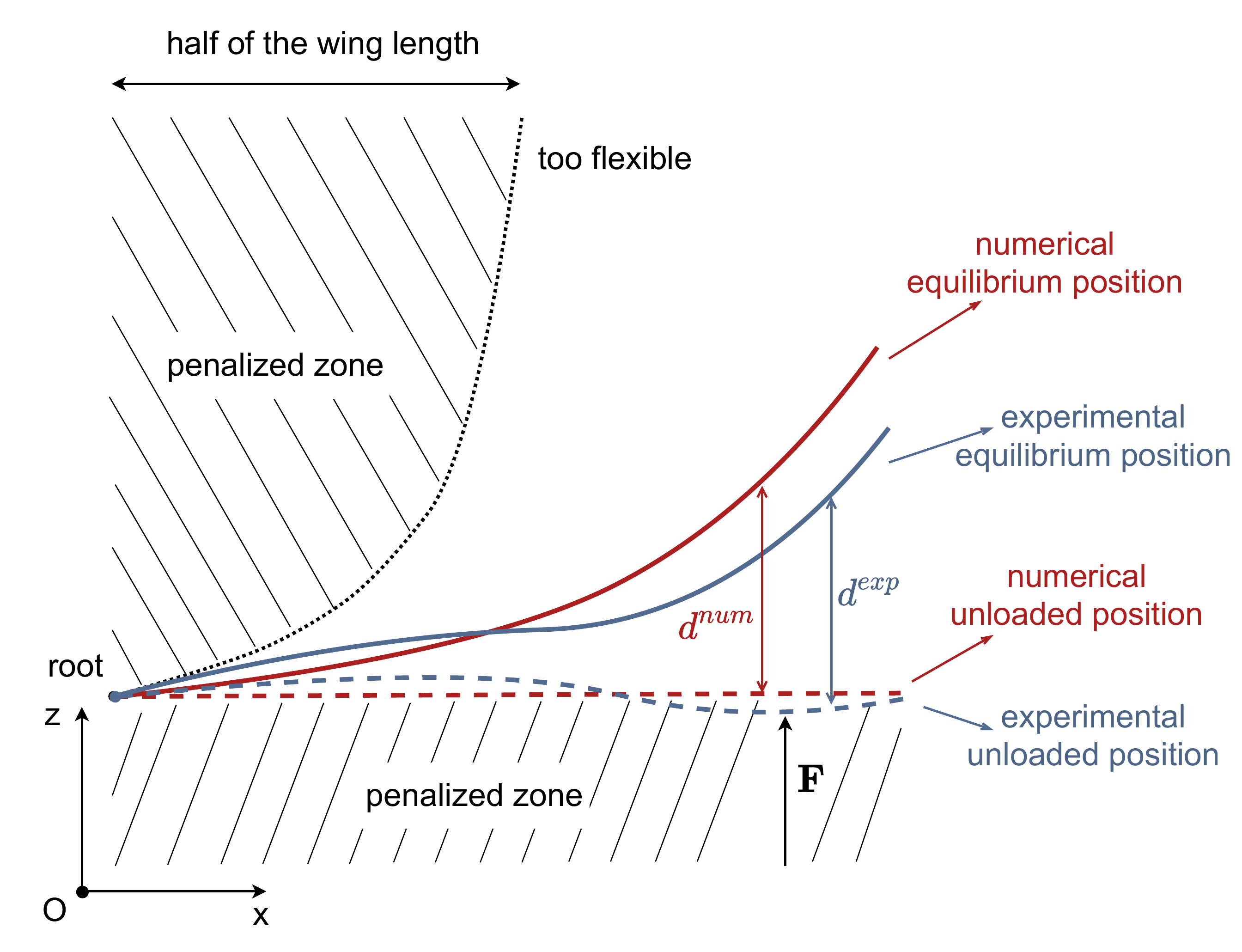}}\\
  \caption{Definition of the cost function for training. Shown is a 2D illustration along the span while actual computations are done in 3D. The difference between the deformation of the experimental wing $d^\mathrm{exp}$ and the deformation of the numerical wing $d^\mathrm{num}$ is used for the calculation of the cost function in eqn.~(\ref{eqn:optimization_objective_function}). The penalized zone below the unloaded wing (the dashed line) corresponds to non-physical solutions while the penalized zone above the dotted line corresponds to the too flexible case.}
  \label{fig:optimization_error_calculation_demonstration}
\end{figure}

For training a numerical wing model, the Euclidean distance ($L^2$-norm) between the nodes of the reference model at the position $\mathbf{x}_i^{r}$ and the learning data at the position $\mathbf{x}_i^{o}$ is usually defined as cost function~\cite{MSMLloyd}. However, this way of defining the cost function based on the equilibrium vertical heights of these nodes can cause some errors. Our numerical wing is flat when unloaded but the real blowfly wing has corrugation and camber which lead to nonzero error even at the initial state when both wings are at rest. To exclude this error, the cost function is instead calculated based on the wing deformation, which measures the difference between the initial and the equilibrium height (cf.~Fig.~\ref{fig:optimization_error_calculation_demonstration}). Moreover, the displacements of each wing were measured $N_\mathrm{exp}$ times, for each point force (Fig.~\ref{fig1meta}D). As a result, a single cost function $h$ is calculated for an individual wing including all $N_\mathrm{exp}$ measurements,
\begin{equation}
    h(\boldsymbol{\kappa}) = \sqrt{ \sum_{j=1}^{N_\mathrm{exp}} \frac{\sum_{i=1}^N \left( d_{i,j}^\mathrm{exp}(\boldsymbol{\kappa}) - d_{i,j}^\mathrm{num}(\boldsymbol{\kappa}) \right) ^2}{N_j}}
\label{eqn:optimization_objective_function}
\end{equation}
where $N$ is the number of data points on the wing, $d_i^\mathrm{exp}$ and $d_i^\mathrm{num}$ are the deformation of these points belonging to the reference model $\mathbf{x}_i^\mathrm{exp}$ and the learning model $\mathbf{x}_i^\mathrm{num}$, respectively.

Because the mass points are defined using a triangular Lagrangian grid, the vertical heights at grid nodes of the Eulerian grid, projected onto the modeled wing, need to be interpolated. We consider an Eulerian grid node $\mathbf{x}$ and a triangular element with three Lagrangian vertices $\mathbf{X}_i$, $\mathbf{X}_j$ and $\mathbf{X}_k$ whose heights are $z_i$, $z_j$ and $z_k$, respectively. When the Eulerian point is projected onto the triangle, there are two possibilities:

\begin{itemize}
\item The projection of $\mathbf{x}$ is outside the triangle and the interpolation cannot be done;

\item The projection of $\mathbf{x}$ is inside the triangle or on one of its three edges. Then, the height of the projection of $\mathbf{x}$ is calculated by using barycentric interpolation:
\begin{equation*}
z = \frac{A_i}{A}z_i + \frac{A_j}{A}z_j + \frac{A_k}{A}z_k 
\end{equation*}
where $A=\mathrm{Area}(\mathbf{X}_i,\mathbf{X}_j, \mathbf{X}_k)$, 
$A_i=\mathrm{Area}(\mathbf{x},\mathbf{X}_j,\mathbf{X}_k)$,
$A_j=\mathrm{Area}(\mathbf{x},\mathbf{X}_k,\mathbf{X}_i)$ and 
$A_k=\mathrm{Area}(\mathbf{x},\mathbf{X}_i,\mathbf{X}_j)$.

\end{itemize}

Based on the nature of the problem, two penalized zones are specified as shown in Fig.~\ref{fig:optimization_error_calculation_demonstration}. A point force is applied to the wing from below. Thus, any solution giving an equilibrium position below the dashed line will be judged non-physical. On the other hand, if the horizontal distance from wing root to tip is smaller than $R/2$, the wing is considered too flexible. In these cases, the cost function will be assigned a large value in order to exclude these sets of parameters.
    
\subsubsection{Covariance matrix adaptation evolution strategy:\label{subsec:CMAES}}

The algorithm used for optimizing the cost function $h(\boldsymbol{\kappa})$ in eqn.~(\ref{eqn:optimization_objective_function}) has to be chosen carefully. A difficulty is the existence of local minima in which an optimization algorithm can get trapped and classical methods seem to be ineffective. Amongst alternative methods, genetic algorithms have appeared as good solutions due to their ability to deal with complex optimisation problems and parallelism. In related work, Nogami et al.~\cite{MSMNogami} estimated stiffness and damping coefficients of several mass-spring-damping models using genetic algorithms. Bianchi et al.~\cite{MSMBianchiParam,MSMBianchiMesh} proposed using genetic algorithms to optimize the stiffness values 
together with the mesh topology. Louchet et al.~\cite{MembraneLouchet} used evolutionary algorithms to identify the parameters of a physical model of fabrics. Joukhadar et al.~\cite{MSMJoukhadar} optimized the physical parameters of a masses/springs based system such as elasticity, viscosity, plasticity with a genetic algorithm based approach. In this contribution, we propose an approach to determine the spring constants of the mass-spring model by using an evolution strategy.

The CMA-ES (Covariance Matrix Adaptation Evolution Strategy) is an optimization algorithm based on the process of natural selection where the most well-suited individuals are selected for reproduction of the next generation. The method is developed for complex non-linear non-convex black-box optimisation problems in continuous domain \cite{GenOpHan01,GenOpHan03}, especially in cases where an analytical formulation of the cost function cannot be easily derived. In other words, function values at search points are the only accessible information on the cost function $h$.

A standard CMA-ES, as described in detail in \cite{GenOpHan01,GenOpHan03}, is used with weighted intermediate recombination, step size adaptation, and a combination of rank-$\mu$ update and rank-one update. The algorithm addresses the following optimization problem: minimize a nonlinear multivariable cost function from search space $\mathcal{S} \subseteq \mathbb{R}^n_+$ to $\mathbb{R_+}$. Let $x_k^{(g)}$ be the $k^{th}$ offspring (solution candidate) at the generation $g$ (iteration). The new offsprings at the next generation $g+1$ are given by:
\begin{equation}
    \mathbf{x}_k^{(g+1)} = \mathbf{m}^{(g)} + \boldsymbol{\sigma}^{(g)} \mathcal{N} (\mathbf{0},\mathbf{C}^{(g)}) \ \ \textrm{for} \ \ k=1...\lambda
\label{eqn:CMAES_update_next_generation}
\end{equation}
where $\boldsymbol{\sigma}^{(g)}$ is the overall standard deviation (step size), $\mathcal{N}$ denotes the normal distribution with zero mean and $\mathbf{C}^{(g)}$ is the covariance matrix. After each iteration, the offspring are evaluated on the cost function $h$ and sorted in decreasing order as:
\begin{equation*}    \{x_{i:\lambda} \mid i=1\dots\lambda\} = \{x_i\mid i=1\dots\lambda\}        
\end{equation*}
and
\begin{equation*}    
     h(x_{1:\lambda})\le\dots\le h(x_{\mu:\lambda}) \le \cdots \le h(x_{\lambda:\lambda}),       
\end{equation*}
Only the best-suited $\mu$ candidates are chosen as the parents for the reproduction of the next generation. Here, $\mathbf{m}^{(g)}$ is the mean of the sampling distribution which is the weighted intermediate recombination of the $\mu$ best candidates from the previous generation:

\begin{equation}
    \mathbf{m}^{(g)} = \sum_{i=1}^{\mu} w_i x_i^{(g)}
\end{equation}

A super-linear relation is used for the recombination, given by:

\begin{equation}
    w_i = \frac{\ln{(\mu + 1)} - \ln{(i)}}{\sum_{i=1}^{\mu} \left( \ln{(\mu + 1)} - \ln{(i)} \right) }
\end{equation}

The second term on the right hand side of eqn.~(\ref{eqn:CMAES_update_next_generation}) is a normally distributed random vector which represents the mutation of the evolutionary strategy. It is obvious to see that the parameters of the normal distribution play an important role in the performance of the optimization. At each iteration, the step size $\boldsymbol{\sigma}^{(g)}$ and the covariance matrix $\mathbf{C}^{(g)}$ are updated in a way that will increase the probability of producing the best offspring for the next generation.  In short, the CMA-ES algorithm implements a principal component analysis of the previously selected mutation steps to determine the new mutation distribution. Due to long and complicated formulae, we refer readers to \cite{GenOpHan01,GenOpHan03} for more details as well as the mathematical derivation of the covariance matrix $\mathbf{C}$. The CMA-ES does not require manual parameter tuning for its application. In fact, the choice of strategy internal parameters is not left to the user (arguably with the exception of population size $\lambda$). Finding good strategy parameters is considered as part of the algorithm design, and not part of its application.

\subsubsection{Numerical setup of the training process:}

Since the evaluation of the cost function is expensive, the code is run in parallel using the Message Passing Interface (MPI) where each evaluation of the cost function is a unique MPI process that is mapped onto the available cores.

To validate our training approach, we first exclusively use numerical data as a reference. Given a set of parameters $\kappa \in \mathbb{R}^n_+$, numerical simulations of the static bending experiment in section~\ref{sec:measurements} are performed using the numerical model (Fig.~\ref{fig:validation}A), with the same forces and application points as in the subsequent training with experimental data. Starting from a random $\kappa_0$, we then perform training with this data (Fig.~\ref{fig:validation}B), thus verify if the training recovers $\kappa$ from this numerical data.

By default, the population size for CMA-ES is $n_\mathrm{pop}=3\ln(n)+4$ \cite{pCMALib}, which yields $n_\mathrm{pop}=12$. However, taking into account the number of available CPU, $n_\mathrm{pop}=16$ was chosen for training, which may slightly improve convergence. Consequently, we ran the validation on $16$ CPUs for $2500$ CPU hours.

In addition, the search space for our problem was restricted using the two penalized zones shown in Fig.~\ref{fig:optimization_error_calculation_demonstration}. Some tests had been performed quickly, before the CMA-ES algorithm was employed, to determine which value of the Young's modulus $E$ would give us an equilibrium position in these two zones. The upper bound for the Young's modulus was \SI{100}{\giga\pascal} because when $E$ was greater than this value, the wing was too stiff and remained almost undeformed under the applied forces. On the other hand, the lower bound for the Young's modulus was set at \SI{0.1}{\giga\pascal} since smaller value of $E$ resulted in an equilibrium in the too-flexible zone. Then based on the relation between the Young's modulus and the bending spring stiffness, the bounds for the joint stiffness were estimated at \SI{0.1}{\newton\centi\metre\per\radian} for the lower bound and \SI{1000}{\newton\centi\metre\per\radian} for the upper bound. These constraints then allow us to speed up the searching process. 

Two stopping criteria were set to determine when to end the searching process. The first one was when the cost function $h$ is smaller than a fixed value. Ideally, it would be zero if we had exactly the same deformation of both wings. This was however not probable in practice and we determined the stopping criterion for the cost function based on the validation test. When the cost function given by eqn.~(\ref{eqn:optimization_objective_function}) was smaller than $10^{-4}$, the difference between two wings is considered negligible. The second criterion was the maximum number of iterations performed by the algorithm. Since the maximum wall time for running the optimization was limited by the supercomputer, the maximum number of iterations was set to $100$ to prevent the computing time exceeding this restriction. This corresponds to more than $4250$ CPU hours for each run.

The wing stiffness of nine individuals were then optimized by this algorithm. To speed up the convergence, each optimization was run with a larger population size of $64$. Based on a sensitivity analysis, we found the Young's modulus $E$ to be the most important parameter. Consequently, we first optimised 11 parameters, corresponding to the Young's modulus $E$ and the stiffness of 10 joints, then $E$ was fixed and only the stiffness values of the 10 joints were optimized for the second run. This procedure led to 2 optimization runs for each individual where each run took more than $4250$ CPU hours.

\subsection{Coupling between the Wing Model and the Fluid Solver for Fluid-Structure Interaction Simulations}

In order to investigate the aerodynamic performance of the optimized wings, we developed a solver for simulation of fluid--solid interaction problems. We integrate the mass-spring model with the incompressible Navier--Stokes solver FLUSI~\footnote{https://github.com/pseudospectators/FLUSI}, details can be found in \cite{Flusi}. FLUSI is a parallel solver based on the Fourier pseudo-spectral method, which resolves all spatial scales of the vortical flow about the flapping wings. The no-slip boundary condition is imposed on the wing surfaces using the volume penalization method \cite{VolPenaAngot}. To construct the volume penalization mask function for the fluid forcing, the distance function is computed by cycling over all the triangles of the Lagrangian grid. The mask function is assigned the value of the regularized two-sided step function of the minimum distance. To transmit forces from the fluid to the solid, we delta-interpolate the pressure near the external boundary of the mask function and calculate the pressure differential across the wing. In the relevant range of the Reynolds numbers (75-4000), wing deformation is caused mainly by the static pressure and the viscous fluid tension is considered negligible~\cite{DickinsonViscous1,DickinsonViscous2,PreidikmanViscous3}. For time-stepping, the coupled fluid-solid system is advanced by employing a semi-implicit staggered scheme, as explained in \cite{HungFlappingBB_TAML}. 
On the one hand, we advance the fluid by using the second order Adams--Bashforth (AB2) scheme. On the other hand, the second-order backward differentiation formula (BDF2) is used in the solid solver. The two modules are weakly coupled such that the solid solver, at a given time step, uses the pressure differential computed at the previous state of the solid model. The net fluid-dynamic forces and torques acting on the wings are evaluated by volume integration of the penalization term \cite{VolPenaDmitry}. 
For further details on the numerical methods including validation we refer to \cite{Flusi, HungRevol, HungFlappingBB_TAML}.

Fully coupled simulations are then carried out using the setup shown in Fig.~\ref{fig1meta}E. From an aerodynamic point of view, the insect body acts as a source of drag in a tethered flight context, which explains why we neglect it in this study and simulate the two wings alone. The computational domain is $36 \times 36 \times \SI{18}{\milli\metre}$ large and discretized equidistantly using $1024 \times 1024 \times 512$ grid points, yielding a total of $537$ million grid points (spacing, $\Delta x=$\SI{35.19}{\micro\metre}). Due to the constraint of the volume penalization method, the wing thickness must be at least 4 grid points. Hence, wing thickness in our numerical study is constant and equals $4\Delta x$ corresponding to \SI{140.76}{\micro\metre}. Although it is thicker than the ones found in nature (\SIrange{1}{10}{\micro\meter}~\cite{ReesWingThickness}), the convergence study in~\cite{EKSFLS19} showed that our numerical scheme preserved its accuracy in the limit of thin wings. We consider a tethered problem, i.e., the wing hinges, located at $\mathbf{x}_{pivot,l} = (18,22.5,9)^T \si{\milli\metre}$ and $\mathbf{x}_{pivot,r} = (18,13.5,9)^T \si{\milli\metre}$ \footnote{$T$ indicates the transposed}, are not moving. The wings are kept far apart to avoid any collision between them. They are exposed to a head wind with the mean flow accounting for the insect's forward velocity $({u}_{\infty},0,0)^T \si{\metre\per\second}$, which corresponds to a typical cruising speed of freely flying blowflies \cite{BomphreyCalspeed,ThomasCalWingCorrugation}.

First, we study the influence of intra-species variability of wing stiffness by comparing the aerodynamic performance of all nine wing models. For better comparison, these models must share geometric, kinematic and dynamic similarity. The first two conditions are clearly satisfied since they have the same wing shape and wing kinematics. The third condition is satisfied if all the simulations have the same Reynolds number $Re$. As the models were tested at forward speed, the Reynolds number $Re$ is calculated based on both cruising speed $u_{\infty}$ and mean wing tip velocity $u_{tip}$ as:

\begin{equation}
  Re = \frac{\left( u_{tip} + u_{\infty} \right) c_m}{\nu_{air}}  
\end{equation}
where $c_m$ is the mean chord length and $\nu_{air}$ is the kinematic viscosity of surrounding flow. Among all the studied individuals, individual number 8 is chosen as the reference whose cruising speed is $u_{\infty}=$ \SI{1.35}{\metre\per\second} and the kinematic viscosity of air is $\nu_{air}=$ \SI{1.568d-5}{\square\metre\per\second}. Because each individual has different wing length, the fluid viscosity as well as the cruising speed of other individuals must be adjusted to match the common Reynolds number. These values are presented in table~\ref{tab:nine_cal_numerical_parameters}.  Therefore, any difference in aerodynamic properties between individuals can be explained by the differences in their wings' flexibility.

\begin{table}[]
\centering
\resizebox{1\linewidth}{!}{%
\begin{tabular}{c c c c c c c}
\hline
Individual & \begin{tabular}[c]{@{}c@{}}Wing length\\ $R (\si{\milli\metre})$\end{tabular} & \begin{tabular}[c]{@{}c@{}}Mean chord\\ $c_m (\si{\milli\metre})$\end{tabular} & \begin{tabular}[c]{@{}c@{}}$u_{tip}$\\  (\si{\metre\per\second})\end{tabular} & 
\begin{tabular}[c]{@{}c@{}}$u_{\infty}$\\  (\si{\metre\per\second})\end{tabular} & \begin{tabular}[c]{@{}c@{}}Kinematic\\ viscosity\\      (\si{\square\metre\per\second})\end{tabular} & \begin{tabular}[c]{@{}c@{}}Reynolds\\ number \\ $Re$\end{tabular} \\ \hline \hline \\ [-1.5ex]
1          & 9.1                                                             & 3.037                                                     & 6.775                                                                            & 1.365    & $1.603 \times 10^{-5}$                                                                                                             &                                                           \\ [0.5ex] \cline{1-6} \\ [-1.5ex]
2          & 9.1                                                             & 3.037                                                     & 6.775                                                                            & 1.365    & $1.603 \times 10^{-5}$                                                                                                             &                                                           \\ [0.5ex] \cline{1-6} \\ [-1.5ex]
3          & 8.7                                                             & 2.903                                                     & 6.478                                                                            & 1.305    & $1.465 \times 10^{-5}$                                                                                                            &                                                           \\ [0.5ex] \cline{1-6} \\ [-1.5ex]
4          & 9.1                                                             & 3.037                                                     & 6.775                                                                            & 1.365    & $1.603 \times 10^{-5}$                                                                                                             &                                                           \\ [0.5ex] \cline{1-6} \\ [-1.5ex]
5          & 8.7                                                             & 2.903                                                     & 6.478                                                                            & 1.305    & $1.465 \times 10^{-5}$                                                                                                             & 1542                                                      \\ [0.5ex] \cline{1-6} \\ [-1.5ex]
6          & 9.2                                                             & 3.070                                                     & 6.850                                                                            & 1.380    & $1.638 \times 10^{-5}$                                                                                                             &                                                           \\ [0.5ex] \cline{1-6} \\ [-1.5ex]
7          & 9.1                                                             & 3.037                                                     & 6.775                                                                            & 1.365    & $1.603 \times 10^{-5}$                                                                                                             &                                                           \\ [0.5ex] \cline{1-6} \\ [-1.5ex]
8          & 9                                                               & 3.004                                                     & 6.701                                                                            & 1.350    & $1.568 \times 10^{-5}$                                                                                                             &                                                           \\ [0.5ex] \cline{1-6} \\ [-1.5ex]
9          & 9.1                                                             & 3.037                                                     & 6.775                                                                            & 1.365    & $1.603 \times 10^{-5}$                                                                                                             &                                                           \\ \cline{1-7}
\end{tabular}}
\caption{The wing length and the mean chord length of nine individuals and the corresponding numerical parameters used for the comparison. Individual number $8$ is chosen as the reference. The kinematic viscosity $\nu_{air} = \SI{1.568d-5}{\square\metre\per\second}$ and the cruising speed $u_{\infty} = \SI{1.5}{\metre\per\second}$ of this individual are real values observed nature. For other individuals, these parameters are scaled based on their wing lengths in order to have the same Reynolds number $Re=1542$. }
\label{tab:nine_cal_numerical_parameters}
\end{table}

In the following, in order to investigate the impact of wing flexibility, individual number 8 with the corresponding numerical parameters (cf.~table~\ref{tab:nine_cal_numerical_parameters}) is compared with rigid wings. 

For the aforementioned setup, a thin vorticity sponge outlet, covering the last 20 grid points in $x$-direction, is used to minimize the upstream influence of the computational domain due to the periodicity inherent to the spectral method. The sponge penalization parameter is $C_{sp}=0.05$ \cite{Flusi}, larger than the permeability $C_{\eta}=2.476 \times 10^{-4}$. By construction, the sponge term is divergence-free and hence does not influence the pressure field \cite{Flusi}.

\section{Results and Discussion\label{ResultsDiscussion}}

\subsection{Validation of Wing Model Training Algorithm}
\begin{figure*}
\centering
\includegraphics[width=152.75mm]{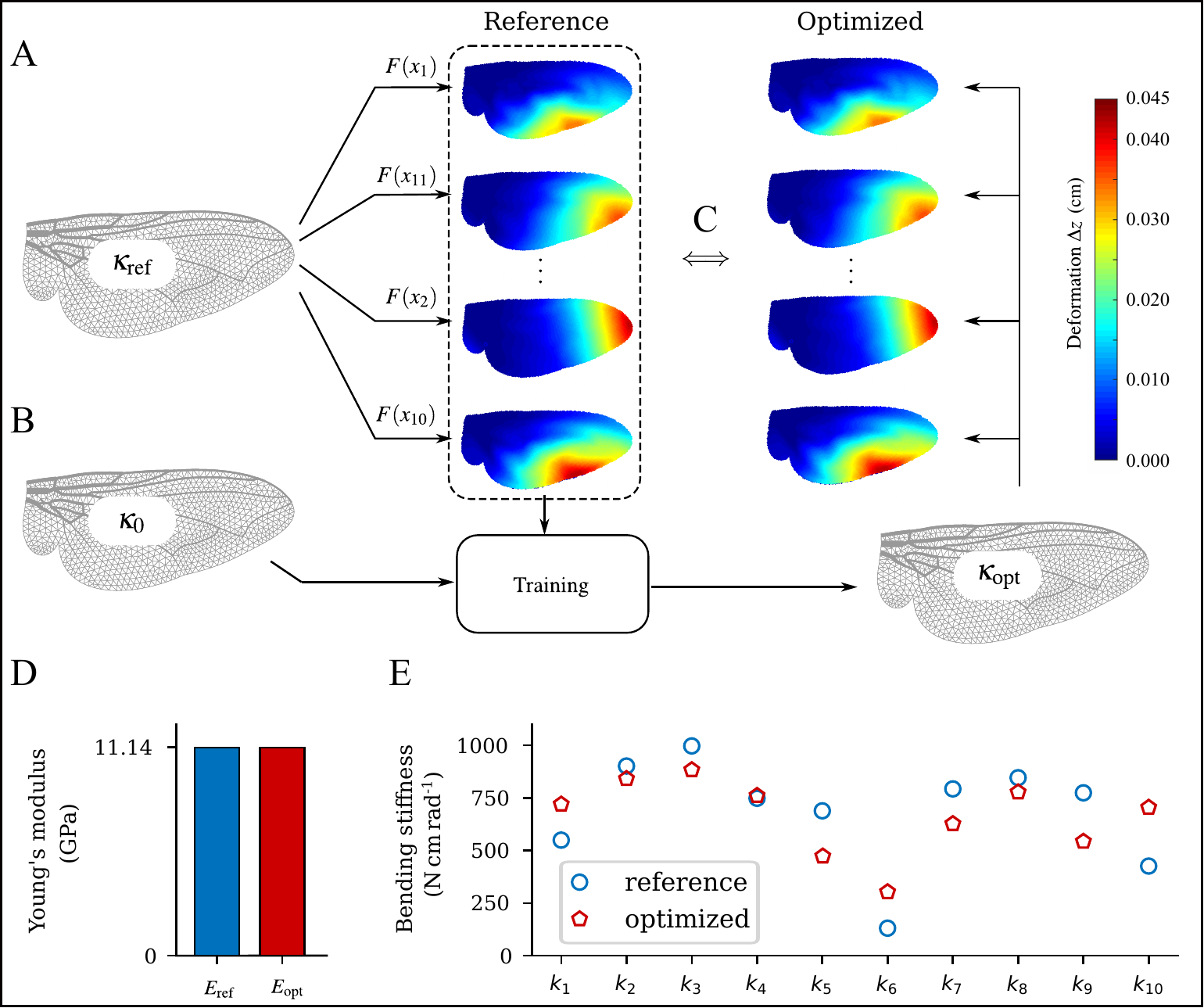}
  \caption{Validation of the training algorithm. (A) Numerical experiments with known parameters $\kappa_\mathrm{ref}$ are used to generate data for training. Only four out of the ten force application points are shown. (B) starting from a random set of parameters $\kappa_0$, the training is done using the numerical reference data, resulting in $\kappa_\mathrm{opt}$. (C) Final deformation fields show good agreement with input data. (D,E) Comparison of Young's modulus and joint stiffness coefficients for reference and optimized data.\label{fig:validation}}
\end{figure*}

First, we validate our training algorithm. After training, the cost function is $h=4.45\cdot 10^{-5}$, and the Young's modulus $E$ (Fig.~\ref{fig:validation}D) is very close to the reference value with a relative error smaller than $10^{-6}$.
This parameter governs the overall deformation of the entire wing and is the most sensitive to the cost function. Thus, it is easier for the algorithm to find the optimal value of $E$. On the other hand, the joints' stiffness values (Fig.~\ref{fig:validation}E) are more difficult to find since each joint solely has effects on the local deformation. By definition, the cost function does not reflect the local deformation of the wing because it averages the Euclidean distance between all the nodes of the reference and the optimized wings.

Figure~\ref{fig:validation}C shows the deformation of the reference- and trained wing in the equilibrium state. Although the difference in joint stiffness coefficients (Fig.~\ref{fig:validation}E) is noticeable, the deformations of the optimizing wing and the reference wing are almost identical. By this criterion, the algorithm is reliable to be used for the optimization.

\subsection{Model Training with Experimental Data}
\label{subsec:resCMA}

\begin{figure*}
\centering
  \includegraphics[width=152.75mm]{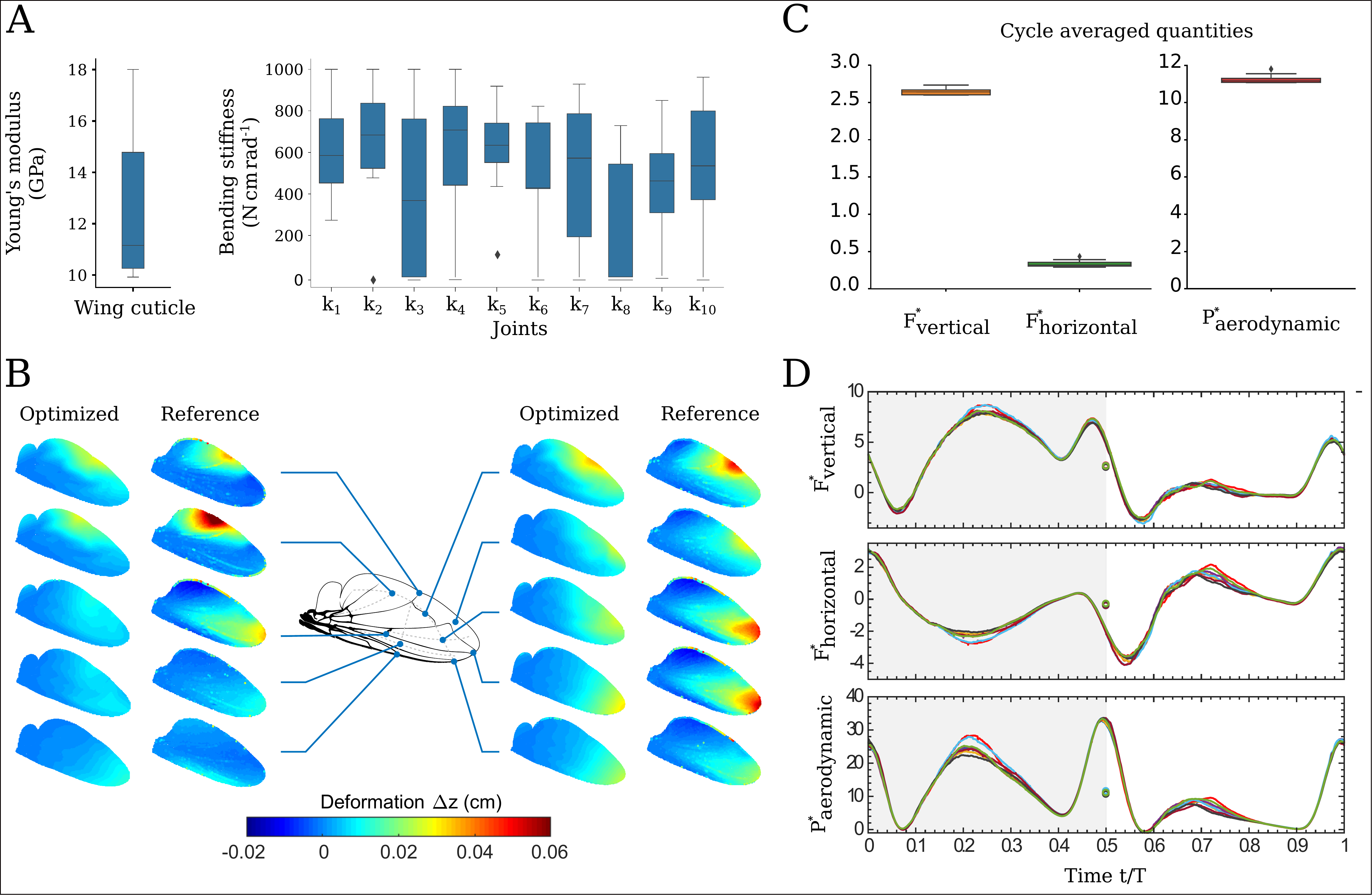}
  \caption{Optimized stiffness and aerodynamic performance of nine studied blowflies. (A) The Young's modulus and the joints' stiffness values of nine individuals optimized by the CMA-ES algorithm. Boxplots are shown with medians and upper and lower quartiles. (B) Training of wing model for individual 8. Shown is the vertical deformation of the numerical (optimized) and experimental (reference) wing with the corresponding force location. (C) Cycle-averaged values over the last two cycles of forces and power calculated for nine individuals. All nine individuals exhibit almost the same aerodynamic behavior, characterized by the small quartiles of the boxplots. (D) Normalized vertical force, horizontal force and aerodynamic power generated by nine individuals with distinct wing flexibility. The time is normalized by the wingbeat period $T=1/f$. Circles represent the cycle-averaged value of forces and power. \label{figmeta_9individuals}}
\end{figure*}

Results of the optimized stiffness parameters are presented in Fig~\ref{figmeta_9individuals}A in the form of boxplots. There is no optimization run succeeding at finding a set of parameters which gives a cost function smaller than $10^{-4}$. All runs were stopped by exceeding the maximum number of iterations at $100$.

 The average value over individuals of Young's modulus is \SI{12.58}{\giga\pascal} with a standard deviation \SI{3.03}{\giga\pascal}. This value is somewhat larger than the value known from previous direct measurements of the Young's modulus of wing cuticle samples, \SI{5}{\giga\pascal} \cite{CuticleProperties,CuticleProperties2}. On the other hand, the joints' stiffness varies significantly among individuals because it depends on various factors such as the distribution of resilin, the shape of veins or the existence of vein spikes \cite{WeisVeinJointStiffness}. Nevertheless, the large spread in parameters can also be explained by non-biological reasons. Firstly, the training process included also the uncertainty of the measurements which cannot be distinguished from the inter-individual differences. Secondly, by definition, the cost function does not reflect the local deformation of the wing because it averages the Euclidean distance between all the nodes of the reference and the optimized wings. Yet, the vein joints solely have effects on the local deformation. 

The deformations of individual number $8$, one of the best solutions with the second smallest cost function, are shown in Fig.~\ref{figmeta_9individuals}B. The left figures represent the deformation measured from experiment while the right figures show the deformation calculated by the wing model. All data are presented in centimeters. In total, 10 measurements corresponding to 10 force locations are plotted.


\subsection{Variance of Wing Stiffness amongst Blowfly Individuals}

In this section, the aerodynamic performance of different individuals with wings of individually tuned stiffness is investigated. We have in total nine sets of stiffness parameters that were optimized in section~\ref{subsec:resCMA}. In the following, all quantities are normalized using the wing length $R$, the wing beat frequency $f$ and the density of air $\varrho_\mathrm{air}=$ \SI{1.225}{\kilo\gram\per\cubic\metre}, unless SI units are explicitly given.

The normalized aerodynamic forces ($F^{*}_\mathrm{vertical}$, $F^{*}_\mathrm{horizontal}$) and the normalized aerodynamic power ($P^{*}_\mathrm{aerodynamic}$) generated by nine individuals in time are presented in Fig.~\ref{figmeta_9individuals}A. Since the wings started at rest and they need some time to stabilize, the data are shown for the $4^{th}$ flapping cycle. In general, the aerodynamic performance of all individuals is almost identical. For each individual, most of the lift (cf.~Fig.~\ref{figmeta_9individuals}A) is generated during the downstroke, followed by a peak and a valley caused by the wing reversal from a downstroke to an upstroke, i.e. the supination. Almost no lift is produced during the upstroke and the wing reversal from an upstroke to a downstroke, or the pronation, produces another peak and valley. The same pattern is observed for the drag and the aerodynamic power. The difference between these individuals is noticeable at the mid-downstroke where the maximum lift generated by individuals 5 and 6 reach $8.8$ while the other peaks are around $8.0$. This amounts to a difference of $10\%$. The corresponding gap between the aerodynamic power at this instant goes up to $22\%$. 

Another way to compare among the individuals is to look at the cycle-averaged values of these quantities. Figure~\ref{figmeta_9individuals}C shows the boxplots of the averaged values calculated for the last two cycles. The ratios between the overall spread, shown by the extreme values at the end of two whiskers, and the median for the three quantities (normalized lift, drag and aerodynamic power) are $5.12\%$, $45.16\%$ and $5.14\%$, respectively. The lengths of the boxes representing the dispersion of the data are small. This indicates the aerodynamic similarity among these individuals even though their wing stiffness vary significantly. 

\subsection{Influence of Wing Flexibility on Aerodynamic Performance of Blowflies}

As shown in the previous section, since the aerodynamic performance varies little among individuals, only individual number 8 is chosen for the next part of our study. The flexible wings will be compared with the rigid wings for studying the influence of wing flexibility on aerodynamic performance of blowflies.

\subsubsection{Aerodynamic performance:}

\begin{figure*}
\centering
  \includegraphics[width=152.75mm]{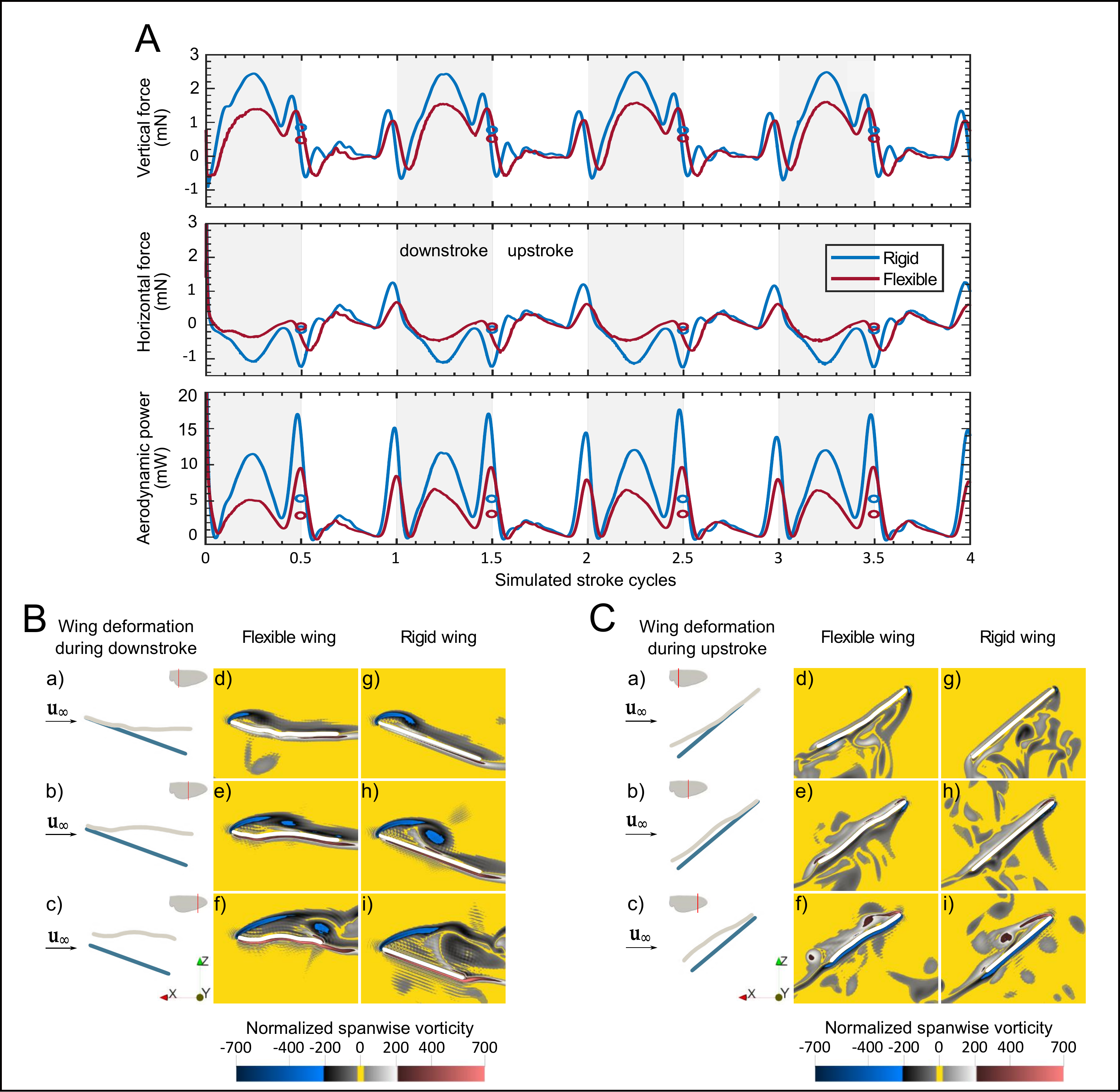}
  \caption{Comparison between the aerodynamic performance of the flexible and rigid blowfly wing models. (A) Time evolution of vertical force and horizontal force generated by the rigid wings (blue) and the flexible wings (red) along with the corresponding aerodynamic power requirement. The time is normalized by the wingbeat period $T=1/f$. Ellipses represent the cycle-averaged value of forces and power. (B,C) Spanwise vorticity, normalized to wing stroke frequency $f$, and wing deformation at the middle of the downstroke ($0.25$ stroke cycle) and the upstroke ($0.75$ stroke cycle), respectively. The visualizations show the cross-section of the left rigid wing (blue) superimposed on the left flexible wing (gray) at $0.25$ (a,d,g), $0.5$ (b,e,h) and $0.75$ (c,f,i) wing length.\label{figmeta_rigid_vs_flexible}}
\end{figure*}

The time history of the vertical and horizontal forces generated by the rigid and the flexible wings, as well as the required aerodynamic power, are shown in Fig.~\ref{figmeta_rigid_vs_flexible}A. Compared to the rigid wings, the stroke reversal, characterized by a peak and a valley of forces, is delayed in the case of flexible wings due to their inertia. Nevertheless, both rigid and flexible wings produce most of the lift during the downward movement and the maximum lift occurs at the middle of the downstroke.

\begin{table}
\centering
\resizebox{1.\linewidth}{!}{%
\begin{tabular}{c c c c c c}
\hline
Wing model  & Lift &  Drag  & Aerodynamic & Lift-to-drag &Lift-to-power \\ [0.5ex] \ \
            &        &        &  power      & ratio & ratio  \\ [0.5ex] \\
            &  (\si{\milli\newton}) & (\si{\milli\newton})  &  (\si{\milli\watt})   &        &        \\ [0.5ex] \hline \hline \\ [-1.5ex]
Rigid       & 0.754   &  0.167  &  5.228 &    4.513 & 0.144  \\ [0.5ex] \hline \\ [-1.5ex]
Flexible    & 0.513   &  0.062  &  3.128 &    8.294 &  0.164 \\ [0.5ex] \hline
\end{tabular}
}
\caption{Cycle-averaged forces and power calculated with the rigid wing model and the flexible wing model. Overall, the rigid wings generate larger forces than their flexible counterparts. The flexible wings have however better performance with higher lift-to-drag and lift-to-power ratios. 

}
\label{table:forces_and_power_Calliphora}
\end{table}

The flexible blowfly wings generate less aerodynamic forces than their rigid counterparts, similar to the result obtained for bumblebee wings studied in~\cite{HungFlappingBB_TAML}. While the two rigid wings produce a maximum lift of \SI{2.49}{\milli\newton}, the one generated by the flexible wings only reaches a maximum value of \SI{1.58}{\milli\newton}. This reduction of $36.55\%$ is due to the wing deformation at the mid-downstroke (0.25 stroke cycle). Figure~\ref{figmeta_rigid_vs_flexible}B (a-c) shows the cross-section of the right rigid wing (blue) superimposed on the right flexible wing (gray) at $0.25$, $0.5$ and $0.75$ wing length. At this instant, the wings move almost in parallel with the oncoming flow and the angle of  attack can be approximated as the angle between the mean flow velocity $\mathbf{u}_{\infty}$ and the wing. As can be seen from the figure, at the proximal (a) and the middle part (b), the leading edge of the flexible wing remains undeformed compared to the one of the rigid wing. The trailing edge is however much more flexible because there are fewer veins to support the membrane. As a result, the trailing edge is pushed upward and adapts its shape to align with the mean flow. This mechanism is caused by the flexibility of the membrane part. It lowers the effective angle of attack, but at the same time reduces the projected area of the wing with respect to the oncoming airflow. The relationship between the effective angle of attack $\alpha$ and the lift-to-drag ratio during translation was derived by Usherwood and Ellington~\cite{UsherwoodLDandAoA} as follows:
\begin{equation}
    \frac{1}{\tan (\alpha)} = \frac{Lift}{Drag} 
\end{equation}
It is then understandable that a lower effective angle of attack results in higher lift-to-drag ratio. Likewise, the required aerodynamic power is smaller because the drag relatively decreases. Characterized by higher lift-to-power ratio ($8.294 > 4.513$) and higher lift-to-drag ratio ($0.164 > 0.144$), flexible wings outperform their rigid counterparts, as presented in table~\ref{table:forces_and_power_Calliphora}. 

Figure~\ref{figmeta_rigid_vs_flexible}B (d-f) shows the spanwise vorticity around the flexible wing while Fig.~\ref{figmeta_rigid_vs_flexible}B (g-i) shows the same kind of data for the rigid wing. From the vorticity distribution in these figures, the development of the leading-edge vortex (LEV) along the spanwise direction of the wing can be seen. For both wing models, the LEV gradually expands toward the wing tip and appears to burst at $75\%$ of the wing length. The size and strength of the LEV generated by the rigid wing are, however, larger than the one created by the flexible wing.

On the other hand, during the upstroke, the dynamic behaviors of both flexible and rigid wings are very similar as little forces are generated. Figure~\ref{figmeta_rigid_vs_flexible}C shows the wing deformation as well as the spanwise vorticity of the left flexible wing and the corresponding rigid wing at the mid-upstroke. The flexible wing almost aligns with the rigid one and the intensities of spanwise vorticity generated by the two models are quite small.

\subsubsection{Pressure distribution and wing deformation:}

\begin{figure*}[th]
\centering
\includegraphics[width=0.65\linewidth]{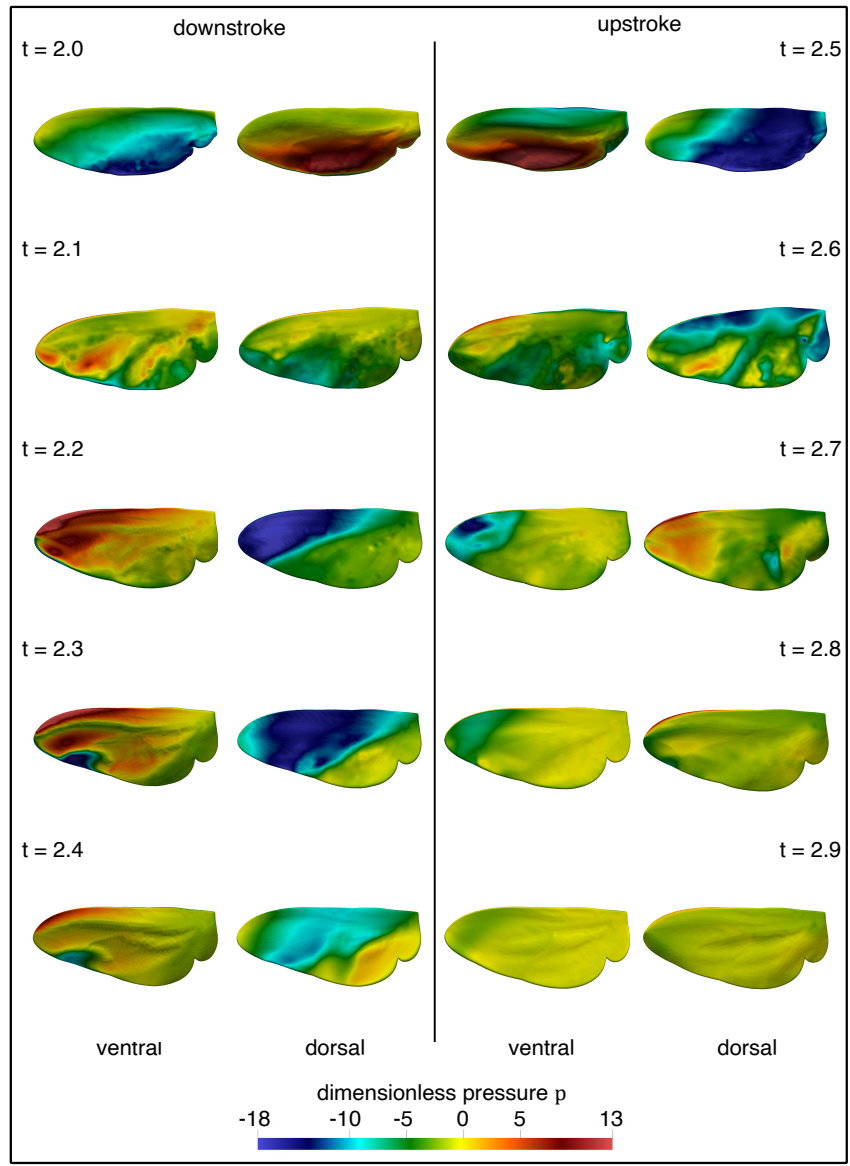}
\caption{Dimensionless pressure distribution on the ventral and dorsal wing side during one flapping cycle. During the downstroke, the difference between the two surfaces is high and generates most of the lift. During the upstroke, the pressure difference is weakened.}
\label{fig:pressure_Calliphora_wing_surface_rainbow_wingsystem}
\end{figure*}

Figure~\ref{fig:pressure_Calliphora_wing_surface_rainbow_wingsystem} presents the dimensionless pressure distribution on the ventral and dorsal wing side. The data are shown for the third cycle.

\begin{figure*}[th]
\centering
\includegraphics[width=0.65\linewidth]{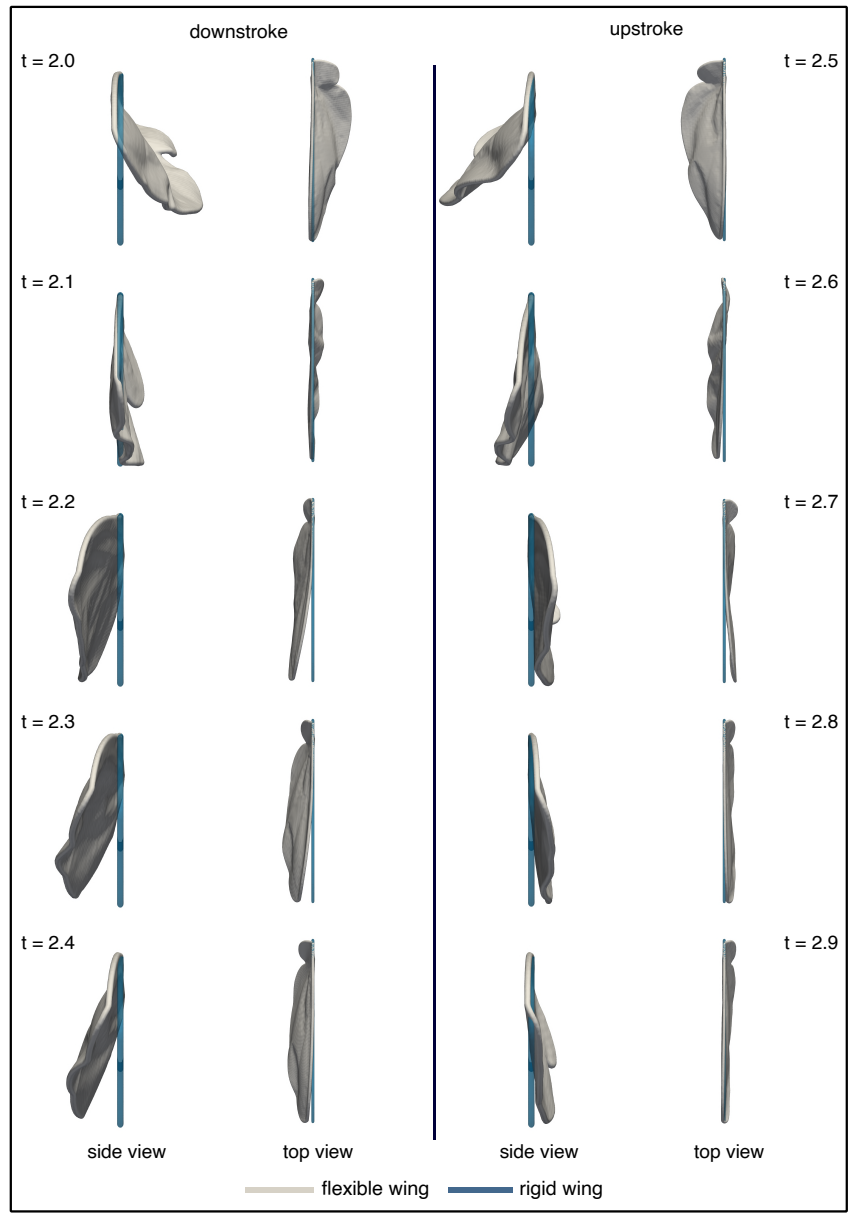}
\caption{Wing deformation during one flapping cycle viewed from the side and the top. The visualizations are shown in the wing system. The wing is deformed significantly in the chordwise direction during the stroke reversals and in the spanwise direction at the middle of the downstroke. The maximum spanwise deflection occurs at the tip of the wing going up to $10^{\circ}$.}
\label{fig:mask_Calliphora_wing_flex_vs_rigid_wingsystem}
\end{figure*}

\begin{figure*}[th]
\centering
\includegraphics[width=0.65\linewidth]{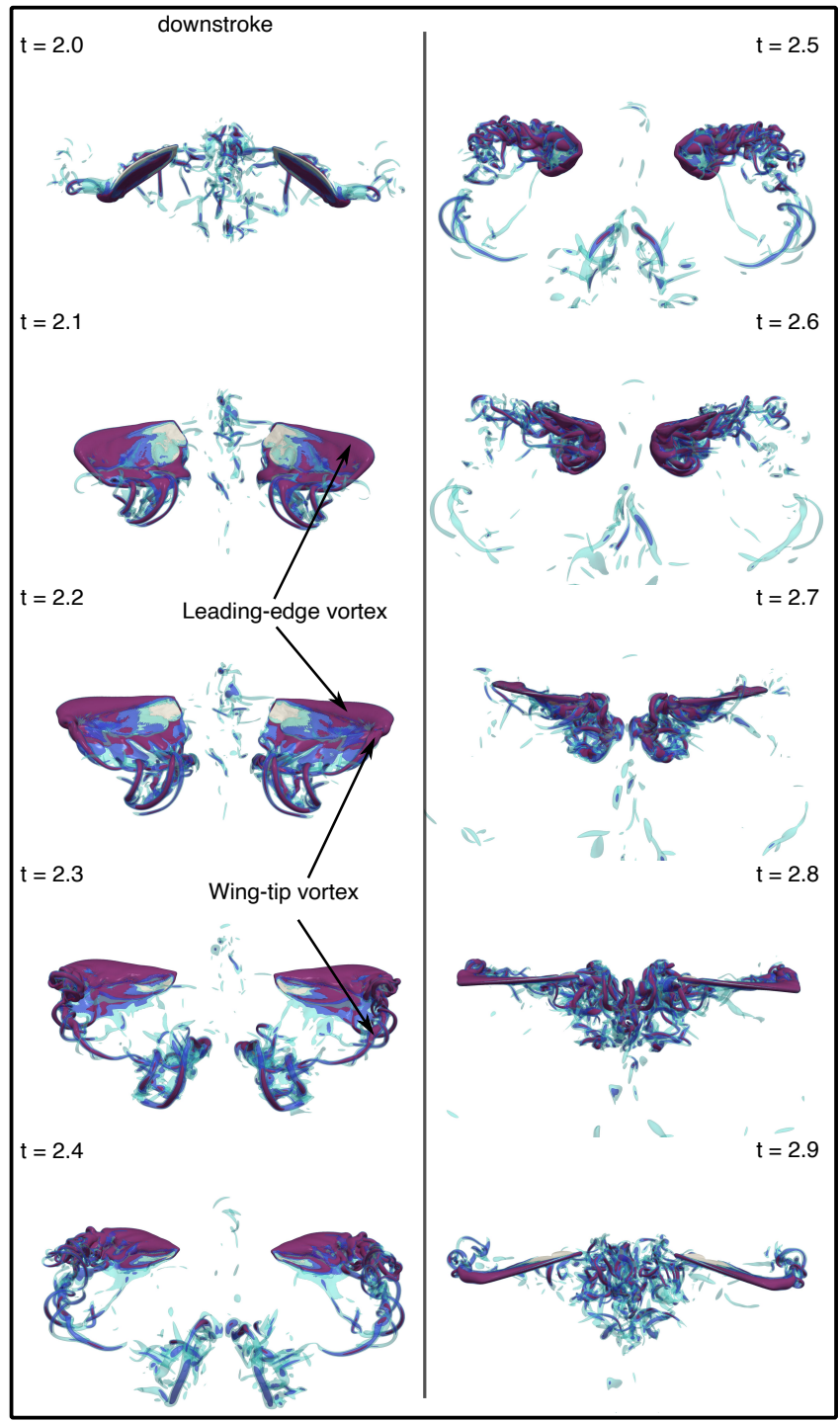}
\caption{Flow structure generated by blowfly wings during flapping. Shown are the iso-surfaces of vorticity magnitude $|\boldsymbol{\omega}|$ of $100$ (light blue), $150$ (dark blue) and $200$ (red). The view point is from the top: the downstroke corresponds to a movement of the wings away from the observer, while the upstroke describes the wings moving toward the observer. }
\label{fig:three_vorabs_Calliphora_wing_Dmitryvapor}
\end{figure*}

During the mid-downstroke ($t=2.2-2.3)$, a large part of the dorsal side becomes a suction zone with low pressure ($p \leq -10$). This low-pressure area is located at the leading edge and expands from the root to the tip. The suction footprint is consistent with the development of the conical LEV observed on the wing surface in Fig.~\ref{fig:three_vorabs_Calliphora_wing_Dmitryvapor}. This finding is anticipated since the existence of LEV on insect wings is associated with the improvement in overall lift production. Figure~\ref{fig:mask_Calliphora_wing_flex_vs_rigid_wingsystem} shows the wing deformation in both chordwise and spanwise directions under the inertial and aerodynamic loading. The deformation in the chordwise direction is understandable since there is no vein to support the membrane belonging to the trailing edge of the wing. The spanwise deformation is, on the other hand, thought-provoking. Combes and Daniels~\cite{CombesI} reported that spanwise flexural stiffness is 1-2 orders of magnitude larger than chordwise flexural stiffness when measuring the forewings of 16 insect species. However, the external forces acting on the wing are strong enough to make the wing deform in both chordwise and spanwise directions. The maximum deflection of the wing leading edge occurs at the wing tip during the mid-downstroke and corresponds to $10\%$ of the wing length. For comparison, Lehmann et al.~\cite{FritzEnergyLoss} observed a maximum wing tip deflection approximately \SI{27}{\degree} at the beginning of the stroke reversal in freely flying blowflies. On the other hand, during the reversals, the wing is deformed only in chordwise direction due to strong inertial force caused by the wing rotation. The dominant contribution of inertial effect to wing torsion at the stroke reversals was earlier pointed out by Ennos~\cite{EnnosInertialForce88,EnnosInertialForce89} for Diptera and agrees with what we observe in our study. Finally, during the upstroke, the pressure difference is weakened since the wings move in the same direction as the mean flow, resulting a low relative oncoming airspeed.

From a 3D point of view, the flow generated by the flexible wings is presented in Fig.~\ref{fig:three_vorabs_Calliphora_wing_Dmitryvapor} for one cycle. The flow structure is visualized by the iso-surfaces of vorticity magnitude $|\boldsymbol{\omega}|$. The development of LEV during the downstroke is considered as the basic aerodynamic mechanism behind the lift production of flapping wings. This spiral structure of the LEV starts to form at the beginning of the downstroke and remains stable until the reversal. The centrifugal force creates a spanwise flow going from the root to the tip which has been explained as the main mechanism helping to stabilize the LEV. However, at approximately three-quarters of the wing length, the LEV starts to detach from the wing surface and forms a wing-tip vortex. During the upstroke, these vortices are weakened and can hardly be seen.

\section{Conclusion}
\label{section:conclusion}

A flexible blowfly wing model has been developed based on the experimental data. The sophisticated structures of the wings were taken into account by distinguishing the vein and the membrane during the meshing procedure. The membrane was modeled as a 2D planar sheet whose tensile strength was much larger than its bending stiffness and the veins were modeled as rods whose bending stiffness values were calculated based on their flexural rigidity $EI$. While the second moment of area $I$ can be estimated using the vein diameters, the Young's modulus $E$ remains somewhat uncertain due to the vast range of known cuticle's property \cite{CuticleProperties}. 

As mechanical properties of insect wings are essential for insect flight aerodynamics, we here presented a numerical method to evaluate the Young's modulus of veins and the joint stiffness of blowfly wings. The mathematical optimization tool CMA-ES~\cite{pCMALib} was employed for determining the right elastic properties by comparing the wing model with static experimental measurements. The method allowed us to find appropriate stiffness values for approximating the static deformation behavior of real insect wings under external point forces. We obtained here nine sets of stiffness parameters for the \textit{Calliphora} wing model.

The high-resolution numerical simulations of a \textit{Calliphora} wing model with the optimized stiffness, flapping in
a moving airflow, allowed us to gain insight into the dynamic behavior of insect
wings, as well as the influence of wing flexibility on the aerodynamic performance
of insects. Firstly, we performed numerical experiments with the full set of stiffness
parameters optimized based on the measurements conducted on nine different individuals.
We found that even though wing stiffness can vary among individuals, their
aerodynamic properties are very similar by comparing dimensionless parameters at
the same Reynolds number. With this conclusion, it is necessary to point out that our findings are restricted to a simple selected wing kinematics pattern. A fly might adapt the wing kinematics according to its wing stiffness. This hypothesis can be tested out by optimizing the wing kinematics based on wing stiffness. However, such studies are computationally expensive or even prohibitive and are left for future work.

We further studied the influence of wing flexibility by comparing between the flexible wings and their rigid counterparts. Under equal prescribed kinematic conditions for rigid and flexible wings, wing flexibility does not enhance lift production but allows better lift-to-drag ratio and lift-to-power ratio. This can simply be explained by changing the effective angle of attack due to wing flexibility. Moreover, from a biological point of view, another benefit can come from the way how forces are distributed throughout the stroke cycle. The decrease of peak force observed during wing rotation helps to reduce stress on muscles and the skeletomuscular system of insects.

In forthcoming work, we will consider detailed numerical investigations of houseflies (\textit{Musca domestica}) for which experimental wing data have been acquired including micro-CT scans of the body.

\section*{Acknowledgments}

Financial support from the Agence Nationale de la Recherche (ANR) (Grant 15-CE40-0019) and Deutsche Forschungsgemeinschaft (DFG) (Grants SE824/26-1 and LE905/17-1), project AIFIT, is gratefully acknowledged. The authors were granted access to the HPC resources of IDRIS under the allocation No. A0102A01664 attributed by Grand Équipement National de Calcul Intensif (GENCI). 
Centre de Calcul Intensif d’Aix-Marseille is acknowledged for granting access to its high performance computing resources.
%
D.K. gratefully acknowledges financial support from the JSPS KAKENHI Grant No. JP18K13693.


\section*{References}
\typeout{}
\bibliography{BB_Optimization_paper_2021}


\end{document}